\documentclass[onecolumn]{aastex6}

\begin{document}


\title{A Study of Central Galaxy Rotation with Stellar Mass and Environment}


\author{Paola~{Oliva-Altamirano}\altaffilmark{1}}
\affil{Centre for Astrophysics \& Supercomputing, Swinburne University of Technology, Hawthorn, VIC 3122, Australia\\ poliva@astro.swin.edu.au}
\and
\author{Sarah~{Brough}\altaffilmark{2}, Kim-Vy~{Tran}\altaffilmark{3}, Jimmy\altaffilmark{3}, Christopher~{Miller}\altaffilmark{4}, Malcom~N.~{Bremer}\altaffilmark{5}, Steven~Phillipps\altaffilmark{5}, Rob~Sharp\altaffilmark{6}, Matthew~{Colless}\altaffilmark{6}, Maritza~A.~{Lara-L\'opez}\altaffilmark{7},  \'Angel~R.~{L\'opez-S\'anchez}\altaffilmark{2,8}, Kevin~{Pimbblet}\altaffilmark{9,10,11}, Prajwal~R.~{Kafle}\altaffilmark{12}, Warrick~J.~{Couch}\altaffilmark{1,2}}
\altaffiltext{1}{Centre for Astrophysics \& Supercomputing, Swinburne University of Technology, Hawthorn, VIC 3122, Australia}
\altaffiltext{2}{Australian Astronomical Observatory, PO Box 915, North Ryde, NSW 1670, Australia}
\altaffiltext{3}{George P. and Cynthia W. Mitchell Institute for Fundamental Physics and Astronomy, Department of Physics and Astronomy, Texas A\&M University, College Station, TX 77843, USA}
\altaffiltext{4}{Department of Physics, University of Michigan, Ann Arbor, MI 48109, USA}
\altaffiltext{5}{H. H. Willis Physics Laboratory, University of Bristol, Tyndall Ave, Bristol, BS8 1TL, UK}
\altaffiltext{6}{Research School of Astronomy \& Astrophysics, Australian National University, Cotter Road, Weston Creek, ACT 2611, Australia}
\altaffiltext{7}{Instituto de Astronom\'ia, Universidad Nacional Aut\'onoma de M\'exico, A.P. 70-264, 04510 M\'exico, D.F., M\'exico}
\altaffiltext{8}{Department of Physics and Astronomy, Macquarie University, NSW 2109, Australia.}
\altaffiltext{9}{School of Physics, Monash University, Clayton, Victoria 3800, Australia}
\altaffiltext{10}{Monash Centre for Astrophysics (MoCA), Monash University, Clayton, Victoria 3800, Australia}
\altaffiltext{11}{Department of Physics and Mathematics, University of Hull, Cottingham Road, Kingston-upon-Hull, HU6 7RX, UK}
\altaffiltext{12}{International Centre for Radio Astronomy Research (ICRAR), The University of Western Australia, 35 Stirling Highway, Crawley, WA 6009, Australia}
\begin{abstract}
We present a pilot analysis of the influence of galaxy stellar mass and cluster environment on the probability of slow rotation in 22 central galaxies at mean redshift $z=0.07$. This includes new integral-field observations of 5 central galaxies selected from the Sloan Digital Sky Survey, observed with the SPIRAL integral-field spectrograph on the Anglo-Australian Telescope. The composite sample presented here spans a wide range of stellar masses, $10.9<$~log(M$_{*}/$M$_{\odot})<12.0$, and are embedded in halos ranging from groups to clusters, $12.9<$~log(M$_{200}/$M$_{\odot})<15.6$. We find a mean probability of slow rotation in our sample of P(SR)$=54\pm7$\%.  Our results show an increasing probability of slow rotation in central galaxies with increasing stellar mass. However, when we examine the dependence of slow rotation on host cluster halo mass we do not see a significant relationship. We also explore the influence of cluster dominance on slow rotation in central galaxies. 
Clusters with low dominance are associated with dynamically younger systems. We find that cluster dominance has no significant effect on the probability of slow rotation in central galaxies.  These results conflict with a paradigm in which halo mass alone predetermines central galaxy properties.  
\end{abstract}




\section{Introduction} \label{sec:intro}

The central galaxies of groups and clusters\footnote{Throughout the paper we make no distinction between groups and clusters to give continuity to the analysis.} are generally (but not always) the brightest galaxies in those systems (BCGs). Central galaxies sit at the extreme end of the galaxy mass distribution. Simulations predict that central galaxies have higher merger rates than less massive early-type galaxies \citep{White_1978,Khochfar_2003,De_Lucia_2007,Oser_2010,Tonini_2012,Laporte_2013}. The main reason for this is their privileged position at the bottom of the gravitational potential of the cluster.  
\\\\
Analysis of stellar kinematics is one of the methods that can be used to examine the merger history of galaxies. This is complementary to analysis of stellar populations \citep[e.g.][]{Brough_2007, Oliva_Altamirano_2015,Greene_2015} or stellar mass growth \citep[e.g.][]{Lidman_2012,Oliva_Altamirano_2014}. Initial analyses of the stellar kinematics of central galaxies used long-slit spectroscopy to quantify the stellar velocity and velocity dispersion profiles \citep[e.g.][]{Fisher_1995,Carter_1999, Brough_2007,Loubser_2008}. Integral Field Spectroscopy (IFS) now allows observations which combine spectral with spatial information. The SAURON \citep{Tim_de_Zeeuw_2002} and ATLAS$^{3D}$ \citep{Cappellari_2011} IFS surveys introduced the $\lambda_{R}$ parameter as a proxy for the specific angular momentum of galaxies and introduced a $\lambda$-based classification to distinguish between fast (high specific angular momentum) and slow (low specific angular momentum) rotators.
\\\\
Several simulations have tried to explain the final angular momentum of a galaxy through its merger history \citep[e.g.][]{Bois_2011,Khochfar_2011,Naab_2014}. They showed that there is not a definitive merger history that would result in a fast or slow-rotating galaxy. The final angular momentum of a galaxy is mostly influenced by the fraction of gas involved in a merger and the spin orientation of the merger candidates. \citet{Naab_2014} predicted that the fraction of stellar mass formed in-situ since $z=2$ appears to be a good indicator of dissipation. Galaxies with an in-situ stellar mass fraction $>18$\% show distinct kinematic features, which are commonly linked to gas-rich mergers. However, the simulations are not yet in full agreement with the observations. Some observations suggest that fast rotators can form through merging quenched spirals \citep[e.g.][]{Cappellari_2011,Cappellari_2013b}. Therefore, the evolution of fast and slow rotators remains an open question. 
\\\\
Central galaxies are the perfect laboratory to test these merger scenarios. They are predicted to have gone through more mergers then other massive early-type galaxies and to have ceased their in-situ star formation at $z>1$ \citep{De_Lucia_2007}. \citet{McGee_2009} found that around 40\% of the cluster galaxy population has been accreted from galaxy groups, and central galaxies play a special role in this assembly. When a cluster merges with another cluster or group, the central galaxy is likely to eventually accrete the most massive galaxies of the in-falling halo \citep[e.g.][]{Balogh_2010,De_Lucia_2012} due to dynamical friction. The majority of central galaxies would be expected to have low angular momentum as they are the end product of the hierarchical scenario \citep{van_Dokkum_2010}. While there have been several IFS observations of the emission lines of central cluster galaxies \citep[e.g.][]{Edwards_2009, Farage_2010, Loubser_2013, Hamer_2016} there are too few IFS observations of the stellar kinematics of central galaxies 
to prove this hypothesis.
\\\\
\citet{Jimmy_2013} studied the stellar kinematics of 10 central galaxies in clusters with log(M$_{200}/$M$_{\odot})>13.6$. They found that while the majority of the galaxies (70\%) are slow rotators, 30\% of the sample are fast rotators. They also found that the current slow/fast rotation of these central galaxies was not correlated with recent minor mergers (less than 0.2~Gyrs ago).  \citet{Veale_2016} have recently presented IFS observations of 41 massive (log(M$_{*}/$M$_{\odot})>11.8$) early-type galaxies from the MASSIVE survey \citep{Ma_2014}. These include 27 brightest halo galaxies that are not all central galaxies (for example the sample includes M49 in Virgo but not M87). They found that 80\% of their sample are slow rotators and that the fraction of slow rotators showed a strong dependence on galaxy brightness.
\\\\
Other existing IFS observations of the stellar kinematics of central galaxies are those dedicated to analyzing the kinematic morphology~--~density relationship in clusters \citep[Virgo, Abell 1689, Coma, Fornax, Abell 85, 168 and 2399;][]{Cappellari_2011,D_Eugenio_2012,Houghton_2013,Scott_2014,Fogarty_2014}. The majority of the central galaxies in these clusters are slow rotators, with the exception of Abell 2399, where the central galaxy is a fast rotator. However, none of these studies examined the lower mass group environment so the effect of the shallower gravitational well of groups, where merging is expected to be more active, is unknown. Therefore, the role of environment on central galaxy rotation is yet to be studied.  
\\\\
The cluster environment can be parametrized by halo mass and also the dominance of the cluster ($\Delta$m$_{1,2}$). This refers to the magnitude gap between the brightest and the second brightest galaxy in a cluster \citep{Tramaine_1977,Loh_2006,Smith_2010}. The dominance has been used together with X-ray luminosities to identify fossil groups, thought to have formed at early epochs \citep{Proctor_2011,Harrison_2012}. Small magnitude gaps ($\Delta$m$_{1,2}<1$) likely indicate that the cluster has recently gone through a cluster-cluster merger \citep[e.g.][]{Smith_2010,Dariush_2010,Coenda_2012,Martel_2014}. These cluster-cluster mergers could potentially affect the angular momentum of the central galaxy. However, due to the small numbers of central galaxies observed to date this has not yet been explored.  
\\\\
\citet{Cappellari_2013b} and \citet{Cappellari_2013c} examined the evolutionary channels of early-type galaxies in the size~--~stellar mass diagram. To analyze different environments they use the ATLAS$^{3D}$ sample of field and Virgo cluster galaxies, adding further observations from the Coma cluster. They suggest a connection between the transition of fast to slow rotation and the galaxy's stellar mass. Around log(M$_{*,\rm crit}/$M$_{\odot}) = 11.3$ they observe a larger fraction of slow rotating galaxies with no horizontal elongation in the velocity maps.  \citet{Veale_2016} add their sample of massive galaxies to the ATLAS$^{3D}$ sample and find that the fraction of slow rotators increases with increasing $K$-band luminosity.
\\\\
It is not clear whether central galaxies behave similarly to the early-type galaxies in the ATLAS$^{3D}$ and MASSIVE samples, or whether they follow a different relationship with stellar mass.  To study the influence of mass on the fast/slow rotation of central galaxies it is necessary to analyse a larger sample than that available to date.  
\\\\
In this paper we analyse the effect of environment and stellar mass on the probability of slow rotation in central galaxies for the first time. We present new IFS observations of 5 central galaxies hosted by galaxy groups, observed with the SPIRAL IFS on the Anglo-Australian Telescope (AAT) and add these to a compilation of IFS observations of 17 central galaxies from the literature. The 5 new observations extend the literature sample to lower 
halo masses ($12.9<$~log(M$_{200}/$M$_{\odot})<14.4$).  The compilation of 22 central galaxies spans a wide range of stellar masses, $10.9 <$~log(M$_{*}/$M$_{\odot})<12.0$ and cluster masses from group to cluster, $12.9<$~log(M$_{200}/$M$_{\odot})<15.6$. We examine the relationship of central galaxy rotation with galaxy stellar mass, cluster halo mass and dominance.  
\\\\
In Section~\ref{sect:observations} we describe our observations and data reduction. The measurements used, including the kinematic classification, are described in Section~\ref{sect:derived}. Results are presented in Section~\ref{sect:results}, and discussed in Section~\ref{sec:d} before conclusions are drawn in Section~\ref{sect:conclusions}. Throughout this paper we assume the cosmology H$_0 = 70$~km~s$^{{-}1}$~Mpc$^{{-}1}$, $\Omega_{\rm M} = 0.3$, $\Omega_\Lambda = 0.7$.
\\
\section{Observations and Data Reduction}
\label{sect:observations} 
\begin{figure}
\begin{center}
\includegraphics[angle=90,width=0.4\linewidth]{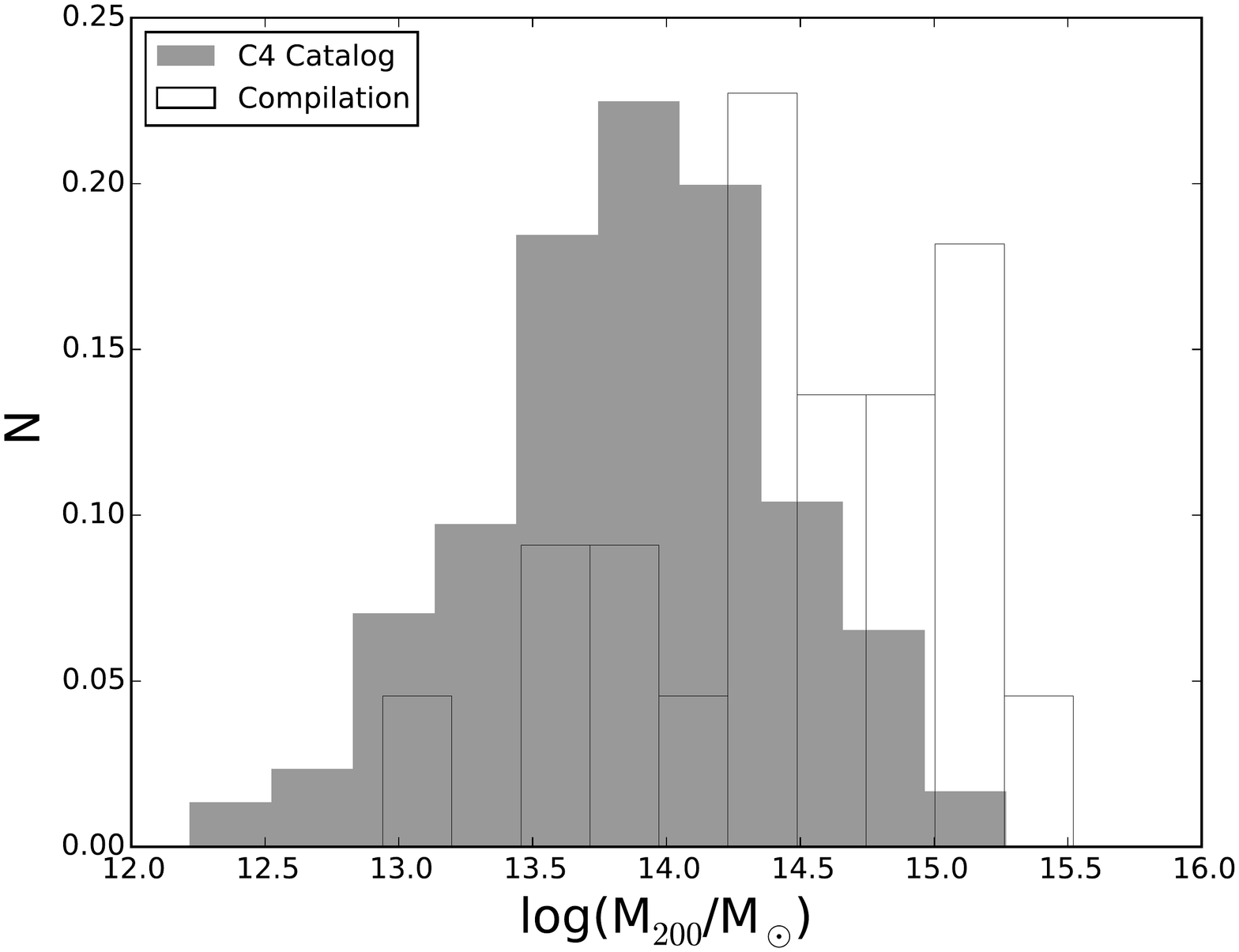}

\includegraphics[angle=90,width=0.4\linewidth]{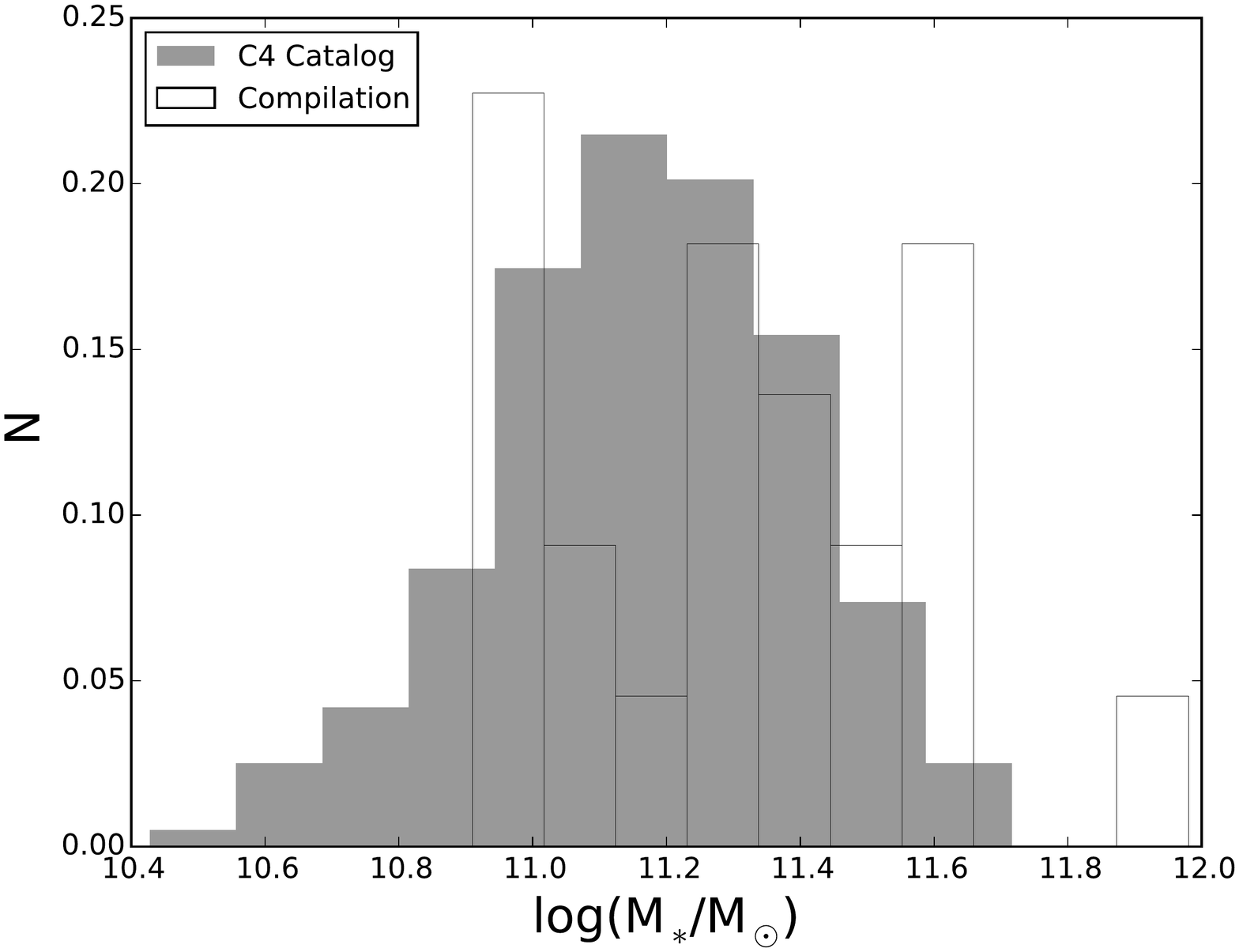}
\caption{\label{fig:vd}Distribution of cluster halo mass (upper panel) and central galaxy stellar mass (lower panel) for the C4 cluster catalog (grey histogram) and the compilation sample of central galaxies presented here (open histogram). The histograms have been normalised so that the total number of galaxies equals 1 for both samples. The compilation sample spans the majority of the halo mass and galaxy stellar mass range of the C4 catalog.%
}
\end{center}
\end{figure}

We are interested in analyzing the effect of environment and stellar mass on the probability of slow rotation in central galaxies. We have therefore compiled the largest possible sample of central galaxies with IFS observations of their stellar kinematics. The compilation sample is made up of new observations of 5 galaxies, 10 galaxies from our previous analysis \citep{Jimmy_2013} and 7 central cluster galaxies from the literature. These are in the Virgo \citep[M87;][]{Cappellari_2011}, Abell 1689 \citep[12;][]{D_Eugenio_2012}, Coma \citep[GMP2921 or NGC 4889;][]{Houghton_2013}, Fornax \citep[NGC 1399;][]{Scott_2014} and Abell 85, 168, and 2399 \citep[019, 042, 086;][]{Fogarty_2015} clusters. The composite sample spans a wide range of stellar masses, $10.9<$~log(M$_{*}/$M$_{\odot})<12.0$, and are embedded in halos ranging from groups to clusters, $12.9<$~log(M$_{200}/$M$_{\odot})<15.6$. We show the cluster halo mass and stellar mass of our compilation sample compared to the central galaxies in the C4 Cluster Catalog \citep{Miller_2005, Von_Der_Linden_2007} from the Third Data Release of the Sloan Digital Sky Survey \citep[SDSS;][]{York_2000} in Figure~\ref{fig:vd}. The two samples are normalized so that the total number of galaxies equals 1 for both samples. While not complete, the compilation sample presented here spans the majority of the cluster mass and central galaxy stellar mass range of the C4 catalog and includes five galaxies in group-mass haloes, log(M$_{200}/$M$_{\odot})<14$. Throughout this paper we define the central galaxy as the brightest galaxy within 0.25 R$_{200}$ of its host group or cluster centroid.  
\\\\
In the following sections we describe the selection and observation of the compilation sample used in this analysis. 

\subsection{New Observations and Data Reduction}\label{sec:ob}
\begin{figure*}
\begin{center}
\hspace*{2cm}
\includegraphics[width=0.75\linewidth]{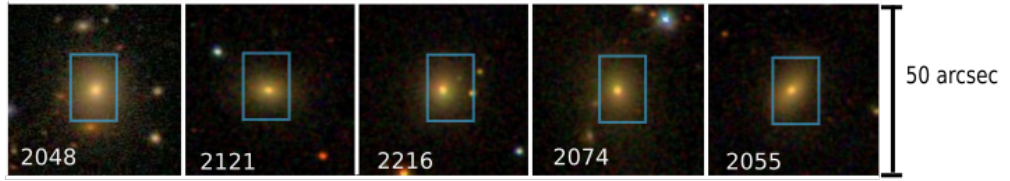}
\caption{\label{fig:sample}SDSS images of the 5 new galaxies observed in this study. The galaxies are sorted by halo mass (from left to right: most massive to least  massive halos). The blue square represents the SPIRAL FOV ($11.2^{\prime\prime}\times22.4^{\prime\prime}$). %
}
\end{center}
\end{figure*}

The new observations presented here are selected from the C4 Cluster Catalog \citep{Miller_2005, Von_Der_Linden_2007}. This catalog includes 625 groups and clusters at $z<0.1$ from the SDSS. The central galaxies in this catalog are defined as the brightest galaxy in the $r-$band closest to the center of the halo \citep{Von_Der_Linden_2007}.  The center of the halo is defined by the position of the galaxy closest to the peak in the density of red-sequence galaxies \citep{Miller_2005}.
The selection criteria for our observations were the following: a) Effective radius between 2$^{\prime\prime}$ and 6$^{\prime\prime}$, to fit within the instrument field-of-view (FOV). b) Redshift range $0.02 < z < 0.2$, to ensure good signal-to-noise (SN). c)~Right Ascension from 175 to 325 deg (J2000). These galaxies are of similar morphology (S\'ersic indices $\sim3.9\pm0.6$).  
To simplify, we identify each galaxy from the C4 catalog by the last 4 numbers of their host cluster i.e. 2121 instead of SDSS-C4-DR3-2121.  
\\\\
The 5 new central galaxies presented here are illustrated in Figure \ref{fig:sample}.  They were observed with the SPIRAL integral-field unit. SPIRAL was an instrument on the 3.9m AAT at the Siding Spring Observatory (NSW, Australia) feeding the AAOmega spectrograph \citep{Sharp_2006}. SPIRAL was composed of 512 fibers arranged in a rectangular array of $32\times16$ with a spatial sampling of $0.7^{\prime\prime} \times 0.7^{\prime\prime}$ per spaxel and a resulting FOV of 22.4$^{\prime\prime}$~$\times$~11.2$^{\prime\prime}$. 
\\\\
These galaxies were observed over 5 dark nights from the 17 to 21st of May 2012. The AAOmega spectrograph has two arms, blue and red, separated by a dichroic beam splitter. We used both the $5700\AA$ and $6700\AA$ dichroics available. While we observed each galaxy with both blue and red arms, in this analysis we focus on data from the blue arm.  Throughout the run we used the low resolution 580V grating which has an average spectral resolution of R$\sim$1900. Depending on the dichroic used, the central wavelength varies from $4800\AA$ to $5700\AA$. The average seeing was $\sim1.1^{\prime\prime}$. Each object was observed in 5 to 7 frames of 2400~seconds each with individual observations dithered by 1-2 spaxels in Right Ascension and Declination in order to avoid the dead elements in SPIRAL. Spectrophotometric standard stars were also observed each night. The details of the observations are summarized in Table \ref{tab:obs}.  
\\\\
\begin{table*}
\begin{center}
\caption{SPIRAL observations. The galaxies are sorted by decreasing halo mass.  
The exposure time shows the number of frames $\times$ seconds of exposure per frame. 
}
\begin{tabular}{ l l c c c c c c}
\hline
      Name & RA & DEC & Observation & Exposure & Dichroic & Seeing\\
     & deg & deg & Date & Time [sec] & [nm] & [$^{\prime\prime}$] \\
    \hline\hline
    2048 & 323.3767 & -8.6407 & 20 May 2012 & 5$\times$2400 & 570 & 1.0\\
    2121 & 314.5097 & -7.5576 & 17 May 2012 & 5$\times$2400 & 570 & 1.0\\
    2216 & 324.0418 &-8.2213 & 21 May 2012 & 5$\times$2400 & 670 & 0.9\\
    2074 & 314.9754 & -7.2607 & 18 May 2012 & 4$\times$2400 & 570 & 0.9\\
    2055 & 316.3945 & -7.5921 & 19 May 2012 & 4$\times$2400 & 570 & 1.3\\
 
    \hline
\label{tab:obs}         
\end{tabular}
\end{center}
\end{table*}

To reduce the raw data we use the AAOmega pipeline, 2\textsc{dfdr}\footnote{\url{http://www.aao.gov.au/science/software/2dfdr/}} \citep{Sharp_2006} version 6.00 (Rockhopper). The process begins by extracting the spectra through tracing the fibers in the flat-field images. These are wavelength-calibrated using arc-lamp frames. The root-mean-square dispersion around the wavelength solution is 0.12\AA. The dispersion around the 5577\AA \  skyline is 0.09\AA. Twilight sky observations are used to provide a relative transmission correction across the IFS, after quartz-halogen flat fields have been used to remove the pixel-to-pixel response variations spectrally. Sky spectra are generated from fibers at the outer edges of SPIRAL not contaminated by galaxy light. Cosmic rays and other spectral defects are rejected with a 2-sigma-clipping mean.
\\\\
Once the individual science spectra have been extracted using 2\textsc{dfdr}, custom \textsc{idl} routines are used to create the science cubes, the `x' and `y' axes contain the spatial coordinates and the `z' axis contains the spectra. We combine all the frames per object by averaging each spaxel (spatial pixel). The bad pixels are set as flux = 0 to avoid any interference with the stacking process. As a result we have one data cube per galaxy with its associated science and variance extensions.

\subsection{Observations from \cite{Jimmy_2013}}
The 10 central galaxies from our previous stellar kinematics analysis \citep{Jimmy_2013} were also selected from the C4 catalog \citep{Miller_2005}. They were selected as having redshifts $0.04<z<0.1$ and 7 were selected because they had close companion galaxies within $\sim10^{\prime\prime}$ while the other 3 galaxies do not have close companion galaxies. 
These central galaxies were observed with the VIMOS \citep{Le_Fevre_2003} IFS on the Very Large Telescope (VLT) in Chile from April to August of 2008 and April to July of 2011. VIMOS was used with the high-resolution blue grism and a spatial sampling of $0.67^{\prime\prime} $pixel$^{-1}$ (BCGs 1153 and 1067 were observed with a spatial sampling of $0.33^{\prime\prime} $pixel$^{-1}$). This gives a field-of-view (FOV) of $27^{\prime\prime}\times27^{\prime\prime}$($13^{\prime\prime}\times13^{\prime\prime}$ for 1153 and 1067). The VIMOS HR blue grism (pre 2012 March 15 version) has a spectral range of $4150-6200\AA$ and a spectral resolution of $0.51\AA$ pixel$^{-1}$. 
The 10 galaxies span a range of stellar masses $10.9<$~log(M$_{*}/$M$_{\odot})<11.7$ and a cluster halo mass range of $13.6<$~log(M$_{200}/$M$_{\odot})<15.3$.

\subsection{Observations from the literature}
We also searched the literature and include prior observations of the stellar kinematics of central galaxies in 7 nearby clusters: Virgo \citep[M87;][]{Cappellari_2011}, Abell 1689 \citep[12;][]{D_Eugenio_2012}, Coma \citep[GMP2921 or NGC 4889;][]{Houghton_2013}, Fornax \citep[NGC 1399;][]{Scott_2014} and Abell 85, 168, and 2399 \citep[019, 042, 086;][]{Fogarty_2014} clusters.  For these clusters we use the same selection criteria as for our observations: the central galaxy is the brightest galaxy within 0.25~R$_{200}$. We summarise the observations as well as our selection of each central galaxy below:

\begin{itemize}
\item{Virgo \citep[M87;][]{Cappellari_2011}: The Virgo cluster at $z\sim0.004$ was observed by the SAURON team. They selected early-type galaxies with $M_K < -21.5$ mag within $D < 42$ Mpc. These galaxies were observed with the SAURON integral field spectrograph on the William Herschel Telescope at the Observatory of El Roque des Los Muchachos on La Palma. All the observations were obtained in the low spatial resolution mode in which the instrument has a FOV of $33^{\prime\prime} \times 41^{\prime\prime}$ sampled with $0.94^{\prime\prime}$ lenslets and with a spectral resolution of $4.2\AA$ full-width at half maximum (FWHM), covering the wavelength range $4800-5380\AA$.  The Virgo cluster has two bright galaxies: M87 and M49. M49 is the brightest in the cluster, however, it is at $\sim0.8$~R$_{200}$. We selected M87 as the central galaxy due to its proximity to the center ($\sim0.0$~R$_{200}$). }
 
 \item{Abell 1689 \citep[12;][]{D_Eugenio_2012}: Abell 1689 is a massive galaxy cluster at $z\sim0.183$. The 30 highest surface brightness early-type galaxies were observed with the FLAMES/GIRAFE spectrograph at the VLT with $R\sim11800$ and rest wavelength range $4858-5521\AA$. Each array of 20 square microlenses has a total FOV of $3.2^{\prime\prime}$ and $0.5^{\prime\prime}$ per lens. We have selected galaxy 12 as the central galaxy by its proximity to the cluster center and its brightness in the $K_s$-band. We note that galaxy 12 is significantly brighter than the 2nd brightest galaxy in the cluster making this an unambiguous decision.}
 
\item{Coma \citep[GMP2921 or NGC 4889][]{Houghton_2013}: Coma is at a redshift $z\sim0.024$. 27 early-type galaxies were selected to match the luminosity function and ellipticity distribution of  a spectroscopically confirmed sample of Coma members. These were observed with the Oxford Short Wavelength IFS (SWIFT; \citealt{Thatte_2006}) image slicer instrument on the 200inch Hale telescope at the Palomar Observatory. SWIFT gives wavelength coverage from $6500 - 10500\AA$ and these observations used the $0.235^{\prime\prime}$ pixel$^{-1}$ spatial scale giving a FOV of $10.3^{\prime\prime} \times 20.9^{\prime\prime}$.   We selected NGC 4889 as the central galaxy by its proximity to the cluster center and its brightness in the $r$-band. }

\item{Fornax \citep[NGC 1399;][]{Scott_2014}: In this study the 10 roundest early-type galaxies with $M_K>-21.5$ mag were observed with the Wide Field Spectrograph (WiFeS; \citealt{Dopita_2010}) IFS on the Australian National University 2.3m telescope. This image-slicing IFS has a FOV of $25^{\prime\prime}\times38^{\prime\prime}$ with $1^{\prime\prime}$ spaxels. The B3000 grating provided wavelength coverage from $3600-5700\AA$ with spectral resolution, $2.2\AA$ FWHM. We have selected NGC 1399 as the central galaxy by its proximity to the cluster center and its brightness in the $K_s$-band. We note that NGC 1399 is significantly brighter than the 2nd brightest galaxy in the cluster making this an unambiguous decision.}

\item{Abell 85, 168, and 2399 \citep[019, 042, 086;][]{Fogarty_2014}. These 3 clusters were observed as part of the Sydney-AAO Multi-IFS (SAMI; \citealt{Croom_2012}) Pilot Survey. The clusters were selected from \citep{Wang_2011}, with $z<0.06$ and Declination$<10^{\circ}$. Galaxies were observed if they were within $10^{\circ}$ of the cluster center and had $M_r<-20.25$ mag. They were observed with the SAMI multi-object IFS mounted on the AAT \citep{Bryant_2015}. Each of SAMI's 13 IFUs (hexabundles) are $\sim15^{\prime\prime}$ in diameter and comprise 61 individual fibres each with a core diameter of $1.6^{\prime\prime}$ arcsec. The SAMI Pilot Survey used the same AAOmega spectrograph with 580V grating in the blue arm as our SPIRAL observations presented above. The central galaxies in Abell 85 and 168 were selected due to their proximity to their cluster centers and their $r$-band brightness. However, Abell 2399 has two dominant galaxies within 0.25~R$_{200}$ \citep[086 and 088;][]{Fogarty_2015} and two brighter galaxies at larger radii \citep[067 and 077;][]{Fogarty_2015}. We select 086 as the central galaxy of Abell 2399, as it is the brightest galaxy within 0.25~R$_{200}$.  }
\end{itemize}

Henceforth, we refer to these heterogeneous observations as `central galaxies in the literature'. 

\subsection{Simulations of Central Galaxies}
Recently, \citet{Martizzi_2014} used a hydrodynamical zoom-in adaptive mesh refinement simulation including Active Galactic Nuclei (AGN) feedback to study central galaxy angular momentum. They found that they could only reproduce the fraction of central galaxy fast rotators observed \citep[][$\sim 30$\%]{Jimmy_2013} if they included AGN feedback. Without AGN feedback, simulations overproduce the fraction of fast-rotating central galaxies. We show these simulated galaxies in the relevant figures throughout the paper. 
\\
\section{Analysis}
\label{sect:derived}
The main purpose of this work is to analyse the influence of stellar mass, and environment (as measured by cluster halo mass and dominance) on the probability of slow rotation in central galaxies. The following sections describe the different measurements used in the analysis. 

\subsection{Photometric measurements: effective radii}
To accurately measure the slow/fast rotation of the galaxies it is essential to know their effective radius, R$_e$. 
We manually calculate the R$_e$ of the 5 newly observed galaxies using the same method as \citet{Kelvin_2012}. We first feed the $r$-band SDSS DR7 images into Source Extractor \citep[Sextractor; ][]{Bertin_1996} to identify the different objects in the image. This provides the central pixel of each object, as well as an empirical R$_e$, magnitude, and orientation of the galaxy. In a second step, we run \textsc{galfit 3} \citep{Peng_2010} on the $r$-band SDSS image using the values from SExtractor as a first guess. We use the -noskyest flag to let \textsc{galfit} calculate the sky level.  The galaxies are fitted with a S{\'e}rsic profile (n$\sim3-6$). To account for contamination from nearby galaxies we use object masking and simultaneous fitting. The accuracy is checked by inspecting the residuals of the fit. 
\\\\

Table \ref{tab:basics} shows the R$_e$ measurements for each galaxy. 

\subsection{Stellar Mass}
For the majority of our sample (19/22 galaxies) we estimate stellar masses using the \citet{Taylor_2011} relationship between $g$- and $i$-band rest-frame colors. \citet{Taylor_2011} derived this empirical relationship by fitting spectral energy distributions to SDSS $ugriz$ imaging assuming a \citet{Chabrier_2003} initial mass function. 
\\\\
We use the Petrosian magnitudes from the NYU Value-Added Galaxy Catalog SDSS DR7 \citep{Blanton_2005}, specifically the k-corrected rest-frame values\footnote{kcorrect/kcorrect.nearest.petro.z0.00.fits; \url{http://cosmo.nyu.edu/blanton/vagc/kcorrect.html}} \citep{Blanton2007}. The galaxies 2048 (from our sample) and 1027 \citep[from][]{Jimmy_2013}, do not have spectroscopic data in the SDSS. We therefore use the redshifts measured in these observations (Section \ref{sec:ppxf}), and extinction-corrected fluxes and k-corrections from the NYU Value-Added Galaxy Catalog to calculate the Petrosian absolute magnitudes and their stellar masses. The final stellar masses come from the following equation  \citep[Eq 8 from][]{Taylor_2011}: 
\begin{equation}
\rm{log}(\text{M}_*/\text{M}_\odot) = 1.15 + 0.70(\textit{g} - \textit{i}) - 0.4 \textit{M} _i
\end{equation}
where \textit{M}$_{\rm i}$ is the absolute $i$ magnitude in AB magnitudes. The 
random uncertainties on M$_*$ derived from this equation \citep[see][]{Taylor_2011} are 0.3 dex. We compared these stellar masses to those presented in \citet{Von_Der_Linden_2007} and find a mean difference log$M_{*,(g-i)}-$log$M_{*,vdL}=0.01\pm0.08$ dex.
\\\\
The central galaxies in the Fornax and Abell 1689 clusters are not observed by the SDSS and the SDSS flux measurement for the central galaxy in Virgo is uncertain due to the large angular size of this galaxy. We therefore calculate stellar masses for these 3 galaxies from the total $K_s$-band magnitudes published in \citet{Scott_2014}, \citet{D_Eugenio_2012} and \citet{Cappellari_2011} respectively, using the relationship from \cite{Cappellari_2013c}: 
\begin{equation}
\rm{log}(\text{M}_*/\text{M}_\odot) = 10.58-0.44(M_{K_s}+23).
\end{equation}
This relationship is an empirical fit to Jeans Anisotropic Multi-Gaussian Expansion modelled masses ($M_{JAM}$) for ATLAS$^{3D}$ early-type galaxies from \cite{Cappellari_2013b} assuming that $M_* =M_{JAM}$.
\\\\
Aperture $K_s$-band magnitudes are available for NGC 4889 and 14/15 of the C4 catalog galaxies from the 2MASS extended source catalog \citep{Jarrett_2000}.  We use these to test the uncertainty between the different methods of stellar mass estimation and find a mean log$M_{*,(g-i)}$ - log$M_{*,K_s} = -0.35\pm0.14$ dex. This is similar to the random uncertainty in the $(g-i)$ estimated stellar masses and the $K_s$-band relationship has been calibrated using total $K_s$-band magnitudes which are not available for the majority of our sample. We therefore use the $(g-i)$ masses for 19/22 galaxies and the $K_s$-band masses for 3/22.  We note that this choice does not affect the conclusions we draw here.
\\\\
Table \ref{tab:basics} lists the stellar masses of all the central galaxies in our sample, highlighting the 3 galaxies with stellar masses derived from $K_s$-band magnitudes.

\subsection{Halo mass}\label{sec:h_mass}
Our sample is composed of central galaxies residing in a wide range of halo masses.  We calculate the halo mass (M$_{200}$) for the whole sample using the cluster velocity dispersion ($\sigma_{cl}$) . M$_{200}$ is the mass contained within R$_{200}$ which is the radius at which the density is 200 times the universal critical density \citep[e.g.][]{White_2001}. We use the following equations to calculate R$_{200}$ and M$_{200}$ from the cluster velocity dispersion \citep[e.g.][]{Finn_2005,Koyama_2010}:
\begin{equation}
\text{R}_{200} = 1.73 \frac{\sigma_{cl}}{1000\text{ km s}^{-1}} \frac{1}{\sqrt{\Omega_\Lambda + \Omega_M (1+z)^3}} \text{Mpc}
\end{equation}
\begin{equation}
\text{M}_{200} = 1.2 \times 10^{15} \left( \frac{\sigma_{cl}}{1000\text{ km s}^{-1}}\right)^3 \frac{1}{\sqrt{\Omega_\Lambda + \Omega_M (1+z)^3}} \text{M}_\odot
\end{equation}

The errors are propagated from $\sigma_{cl}$. 
\\\\
We obtained cluster velocity dispersions for the newly observed galaxies and those from \cite{Jimmy_2013} from the C4 catalog \citep{Miller_2005,Von_Der_Linden_2007}.  The sources of the cluster velocity dispersions for the other clusters are: Coma \citep{Colless_1996}, Abell 1689 \citep{Girardi_1997}, Virgo \citep{Schindler_1999}, Fornax \citep{Drinkwater_2001} and Abell 85, 168 and 2399 (Owers et al., in prep). Two of the clusters (Abell 1689 and Coma) have mass measurements from lensing. The strong-lensing measurement of \cite{Halkola_2006} is within 0.01 dex of the cluster velocity dispersion mass for Abell 1689  while the weak-lensing mass measurement of \cite{Kubo_2007} is 0.23 dex higher than the cluster velocity dispersion measurement for Coma. This is consistent within the uncertainties in the cluster velocity dispersion measurements.
\\\\
The M$_{200}$ for each group and cluster in our sample, and its source, are listed in Table \ref{tab:basics}. 

\subsection{Dominance} 
The clusters from \citet{Jimmy_2013} and the groups and clusters from the new SPIRAL sample do not have X-ray observations, therefore, to track the cluster/group merging status we rely on dominance measurements. The dominance is the magnitude gap ($\Delta$m$_{1,2}$) between the brightest galaxy in the cluster and the second brightest. This measurement has been found to be a good indicator of recent cluster mergers \citep{Tramaine_1977,Loh_2006,Smith_2010}. Statistically, large magnitude gaps are common in halos that have not undergone recent halo-halo mergers. These clusters are homogeneous with strong cool cores, and only a low percentage of the cluster mass resides in cluster substructure \citep[$\sim 3$\%;][]{Smith_2010}. Small magnitude gaps likely indicate a young unrelaxed group and/or a recent cluster merger. These clusters are more heterogenous with large fractions of substructure \citep[$\sim30$\% of the cluster mass;][]{Smith_2010,Dariush_2010,Coenda_2012,Martel_2014}.
\\\\
For the majority of the sample (19/22) we calculate the dominance values using the $r$-band.  For the 14 clusters selected from the SDSS \citep[9 from][5 from our observations]{Jimmy_2013} we use the $r$-band SDSS Petrosian magnitudes to calculate the dominance. However, it is not possible to measure a dominance for the 1050 cluster.
The dominance values for the Coma, and Abell 85, 168 and 2399 and Abell 1689 clusters were calculated using the \textit{r}-band magnitudes found in the literature \citep[][d'Eugenio, private communication]{Houghton_2013,Fogarty_2015}. 
\\\\
The dominance values for the Virgo and Fornax clusters were calculated from the \textit{$K_s$}-band magnitudes in the literature \citep{Cappellari_2011, Scott_2014}.  We expect the dominance values calculated from $K_s$-band magnitudes to be virtually identical to those calculated from $r$-band magnitudes because these galaxies all have old stellar populations ($>6$ Gyr; \citealt{Oliva_Altamirano_2015}) so the light in the $r$- and $K_s$- bands is from essentially the same stars such that ($r-K_s$) is constant \citep[e.g.][]{Smith_2002}. 
We therefore directly compare the $K_s$-band dominance values for Virgo and Fornax to the $r-$band dominance values measured for the rest of the sample. We note that this assumption does not affect the conclusions we draw in this paper.
\\\\
In the Virgo and Abell 2399 clusters the central galaxy is not the brightest galaxy. We are using the cluster dominance to provide a picture of the number of similarly massive galaxies in a cluster. A more mature halo, that has not merged recently, will have fewer similarly massive galaxies and a higher dominance than a halo that has merged recently.  For the two cases (Virgo and Abell 2399) where the central galaxy is not the brightest galaxy we calculate the dominance as $(m_1-m_2)$ even if the central galaxy is not the first or second brightest galaxy.
\\\\
Table \ref{tab:basics} lists the dominance values for the groups and clusters in our sample, the 2 clusters for which $K_s$-band magnitudes were used to calculate the dominance are highlighted.

\subsection{Spectroscopic measurements: stellar kinematics}\label{sec:ppxf}
\begin{figure}
\begin{center}
\includegraphics[width=0.38\columnwidth]{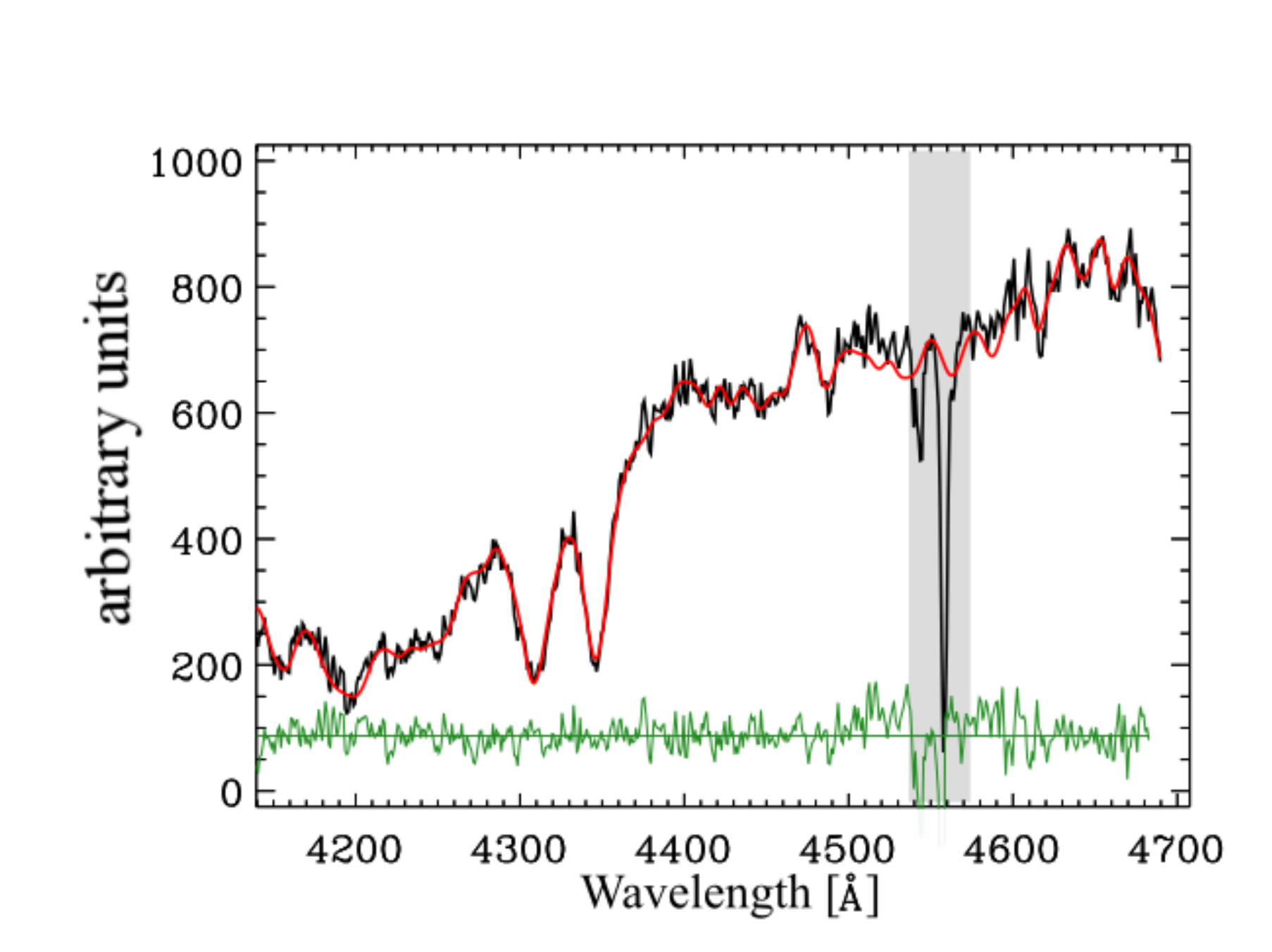}
\caption{\label{fig:spectrum}p\textsc{pxf} fitting of 2048. The black line is the stacked spectra of all the spaxels within 0.4~R$_e$. The red line is the best fit to the observed spectrum. The green line is the residual of the fit. Note that the bad pixels around $\lambda = 4555$~\AA~(grey area) were masked and are ignored by the fit. Overall the stellar template fits the data well ($\sigma = 285\pm7$ km s$^{-1}$, $V=-53\pm6$ km s$^{-1}$, $z=0.094$).%
}
\end{center}
\end{figure}

\begin{figure*}
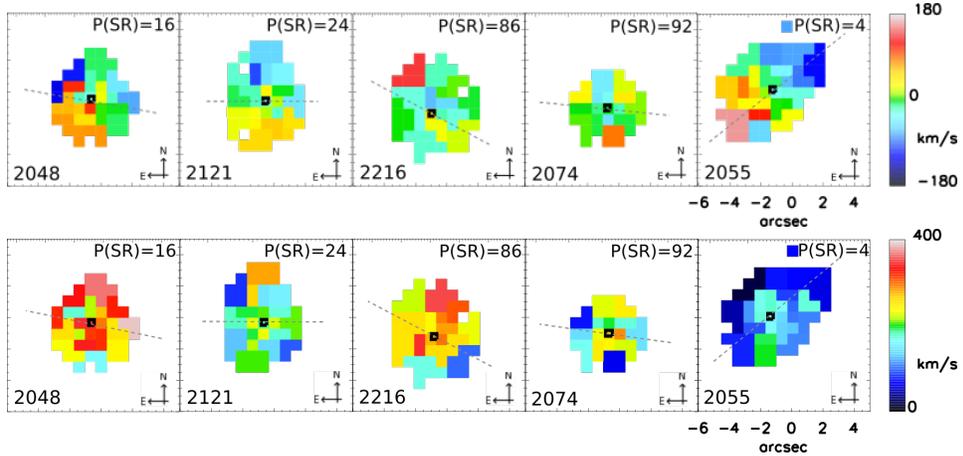

\begin{center}
\hspace*{1.9cm}
\includegraphics[width=0.7\linewidth]{V_maps.pdf}
\hspace*{1.9cm}
\includegraphics[width=0.7\linewidth]{Vd_maps.pdf}

\caption{\label{fig:maps}Kinematic maps of the 5 new central galaxies observed in this study. Velocity is shown in the upper panels. Velocity dispersion is shown in the lower panels. For each galaxy the ID is shown in the lower-left, the probability that the galaxy is a slow rotator is shown in the upper-right and the orientation of the galaxy is shown in the lower-right corner. The dashed line represents the photometric position angle measured by \textsc{galfit}. %
}
\end{center}
\end{figure*}

The stellar kinematics are the key component of this analysis. To ensure a reliable measurement of the stellar kinematic parameters of our new SPIRAL observations: line-of-sight velocity dispersion ($\sigma$) and stellar velocity (\textit{V}), it is necessary to optimize the SN and spatial resolution of the observation. To do so, we use the Voronoi binning method by \citet{Cappellari_2003} which performs a SN cut across all the spaxels (i.e. removes all spaxels with SN below the cut), to later bin the remaining spaxels to reach the minimum SN required. We run Monte Carlo simulations to choose the optimal SN cut and SN required. We find that the optimal SN cut for our data is 3 and the optimal SN after binning is 10. This leaves us with a spatial coverage of 0.2 to 0.6 R$_e$.
\\\\
We measure the stellar kinematics of each binned spaxel with the penalized-PiXel Fitting code \citep[p\textsc{pxf};][]{Cappellari_2004}. p\textsc{pxf} fits the observed spectroscopic data to a library of stellar spectra. In this analysis we use the Medium-resolution Isaac Newton Telescope library of empirical spectra \citep[MILES;][]{Sanchez-Blazquez_2006}. After testing our spectra with all the templates in the MILES library we found that the G6 to M1 stellar templates provided the best fit to the observations as measured by the $\chi^2$ values \citep[as also found by][]{Jimmy_2013}.
The  p\textsc{pxf} fits are performed in the region around the observed strong absorption-line features Ca$_{K+H}\lambda \lambda 3933,3968$ and G-band$\lambda 4307$ (Figure \ref{fig:spectrum}). The resolution of the observed spectra (FWHM = 2.7~\AA) is similar to the resolution of the MILES templates (FWHM = 2.5~\AA). p\textsc{pxf} convolves the library spectra to the FWHM of the galaxy spectra to ensure a better fit. To correct for systemic velocity we use the Vsys flag provided by p\textsc{pxf}.
\\\\
To calculate the uncertainties in the measured velocities and velocity dispersions we use 100 Monte Carlo iterations. In each iteration we add random noise of the order of the residuals from the first fit, to the best fit spectrum. 
The resultant redshifts can be found in Table \ref{tab:basics}. The velocity and velocity dispersion maps are shown in Figure \ref{fig:maps}.

\begin{table*}
\begin{center}
\caption{Galaxy and halo properties. The top section are the new SPIRAL observations. The middle section contains the galaxies from \citet{Jimmy_2013}. The bottom section lists the galaxies taken from the literature. Each section is sorted by decreasing halo mass. (1) Galaxy ID. (2) Cluster ID. (3) Redshift of the galaxy. (4) Effective radius. (5) Stellar mass. (6) Cluster mass. (7) Cluster dominance.}

\begin{tabular}{ l l l l l l l}
    \hline
     Galaxy & Group/  &$z$       & R$_e $ & log M$_*$  & log M$_{200} $ & $\Delta$m$_{1,2}$ \\ 
                & Cluster ID &$\pm0.001$ & [$^{\prime\prime}$] & $\pm 0.10$ [M$_\odot$] & $\pm 0.4$ [M$_\odot$]& [$r$-band] \\ 

    \hline\hline
    2048$^{ }$   & 2048$^{ }$    &0.094$^{ }$ &7.91$^{ }$ & 11.23$^{ }$ & 14.34$^{ }$ & 1.03$^{ }$  \\ 
    2121$^{ }$  & 2121$^{ }$     &0.079$^{ }$ &5.07$^{ }$ & 10.91$^{ }$ & 14.15$^{ }$ & 1.33$^{ }$ \\  
   2216$^{ }$  & 2216$^{ }$     &0.085$^{ }$ &4.75$^{ }$ & 11.01$^{ }$ & 13.76$^{ }$ & 1.43$^{ }$ \\ 
   2074$^{ }$  & 2074$^{ }$     &0.079$^{ }$ &11.42$^{ }$ & 11.07$^{ }$ & 13.71$^{ }$ & 1.22$^{ }$ \\ 
   2055$^{ }$  & 2055$^{ }$     &0.073$^{ }$ &12.27$^{ }$ & 10.93$^{ }$ & 12.94$^{ }$ & 0.02$^{ }$ \\ 
    \hline
    1027$^{ }$ & 1027$^{ }$ &0.090$^{a}$  &6.98$^{a}$ &11.44$^{ }$ & 15.24$^{ }$ & 2.03$^{ }$\\
    1042$^{ }$ & 1042$^{ }$ &0.095$^{a}$  &7.22$^{a}$ &11.47$^{ }$ & 15.01$^{ }$ & 0.36$^{ }$\\
    1048$^{ }$ & 1048$^{ }$ &0.077$^{a}$  &5.17$^{a}$ &11.32$^{ }$ & 14.97$^{ }$  & 1.10$^{ }$\\
    1066$^{ }$ & 1066$^{ }$ &0.084$^{a}$  &5.07$^{a}$  &11.28$^{ }$ & 14.95$^{ }$ & 1.75$^{ }$\\
    2001$^{ }$  & 2001$^{ }$  &0.042$^{a}$  &5.84$^{a}$  &11.38$^{ }$ & 14.75$^{ }$ &0.65$^{ }$\\
    2086$^{ }$ & 2086$^{ }$ &0.084$^{a}$  &4.83$^{a}$  &11.24$^{ }$ & 14.55$^{ }$ & 0.72$^{ }$\\
    1261$^{ }$  & 1261$^{ }$  &0.037$^{a}$  &5.76$^{a}$  &11.01$^{ }$ & 14.37$^{ }$ &1.13$^{ }$\\
    1050$^{ }$  & 1050$^{ }$  &0.072$^{a}$  &8.43$^{a}$&$11.63^{ }$ & 14.35$^{ }$ &---\\
    2039$^{ }$ & 2039$^{ }$ &0.083$^{a}$  &8.82$^{a}$&11.63$^{ }$ &14.33$^{ }$  & 2.17$^{ }$\\
    1153$^{ }$  & 1153$^{ }$  &0.059$^{a}$  &2.39$^{a}$  &10.91$^{ }$ & 13.63$^{ }$ &1.69$^{ }$\\   
    \hline
    12$^{ }$     & Abell 1689   & 0.183$^{b}$ &13.92$^{b}$  & 11.98$^{b}$ & 15.52$^{c}$ & 0.91$^{b}$\\
    NGC 4889 & Coma          & 0.022$^{d}$ &38.00$^{e}$  & 11.30$^{ }$ & 15.20$^{f}$  & 0.22$^{e}$\\
    019$^{ }$   & Abell 85       & 0.055$^{g}$ &16.34$^{g}$  & 11.62$^{ }$  & 15.09$^{h}$ &1.18$^{i}$\\
    086$^{ }$   &  Abell 2399  & 0.058$^{g}$ &4.85$^{g}$    & 11.12$^{ }$  & 14.58$^{h}$ &0.16$^{i}$\\
    042$^{ }$   &  Abell 168    & 0.045$^{g}$ &10.81$^{g}$  & 11.39$^{ }$  & 14.57$^{h}$ & 0.94$^{i}$\\
    M87           & Virgo 	   & 0.004$^{d}$  &81.3$^{j}$    & 11.63$^{j}$  & 14.31$^{k}$   & 0.40$^{j}$\\
    NGC 1399 & Fornax         & 0.005$^{d}$  & 39.9$^{l}$   & 11.50$^{l}$ & 13.80$^{m}$  & 0.51$^{l}$\\
        \hline
        
\end{tabular}
\label{tab:basics}
\end{center}

$^{a}$From \citet{Jimmy_2013}.\\
$^{b}$From \citet{D_Eugenio_2012}. The redshift is for the cluster. The stellar mass was derived from the $K_s$-band magnitude. The $r$-band dominance values were provided by private communication.\\
$^{c}$From \citet{Girardi_1997}. \\
$^{d}$From NASA/IPAC Extragalactic Database.\\
$^{e}$From \citet{Houghton_2013}.\\
$^{f}$From \citet{Colless_1996}.\\
$^{g}$From \citet{Fogarty_2014}. The redshifts are for the clusters.\\
$^{h}$From (Owers et al. in prep.)\\
$^{i}$From \citet{Fogarty_2015}. \\
$^{j}$From  \citet{Cappellari_2011} The stellar mass and dominance were derived from the $K_s$-band magnitudes.\\
$^{k}$From \citet{Schindler_1999}.\\
$^{l}$From \citet{Scott_2014}. The stellar mass and dominance were derived from the $K_s$-band magnitudes.\\
$^{m}$From \citet{Drinkwater_2001}.\\

\end{table*}

\subsection{Specific Angular momentum}
The SAURON \citep{Tim_de_Zeeuw_2002} and ATLAS$^{3D}$ \citep{Cappellari_2011} surveys introduced a new method to quantify the rotation in galaxies. This method is based on the parameter $\lambda_R$, a proxy for the specific angular momentum, and the ellipticity parameter ($\varepsilon$). $\lambda_R$ \citep[][]{Emsellem_2007} is calculated using the stellar velocity (\textit{V}) and line-of-sight velocity dispersion ($\sigma$) as a function of galaxy radius (\textit{R}) as follows:

\begin{equation}
\lambda_R \sim \frac{\langle R|V| \rangle}{\langle R \sqrt{V^2 + \sigma^2} \rangle}
\end{equation}

\begin{figure*}
\begin{center}
\includegraphics[width=1\linewidth]{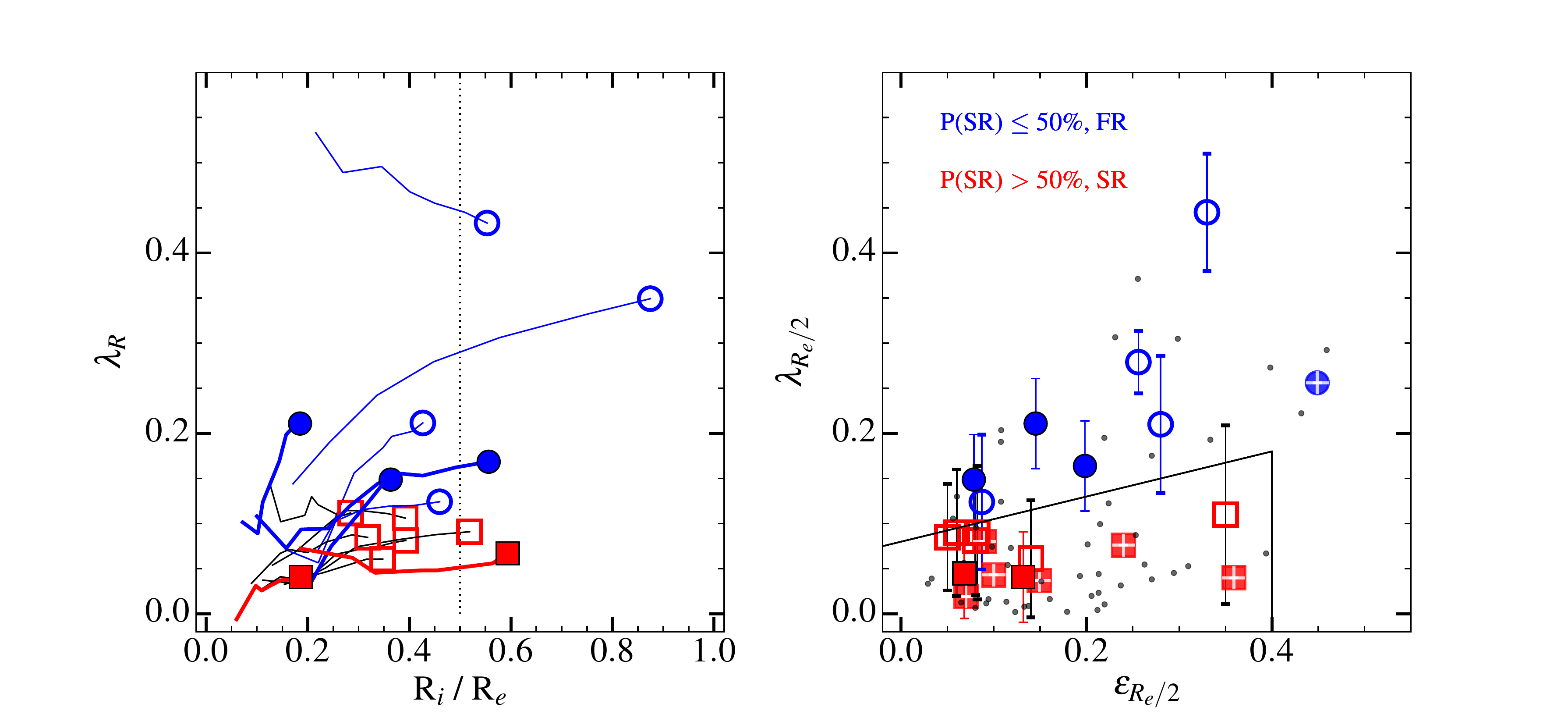}
\caption{\label{fig:rotation}Kinematic classification of central galaxies. The colors represent the central galaxy probability of being a slow rotator: blue circles represent galaxies with P(SR)~$\leq50$\% (classified as fast rotators) and the red squares represent galaxies with P(SR)~$>50$\% (classified as slow rotators). The newly observed central galaxies are shown as filled symbols, the central galaxies from \citet[][]{Jimmy_2013} as open symbols, and the central galaxies from the literature as crossed symbols.  Left-hand panel: $\lambda_R$ vs radius. The vertical dotted line shows the fiducial radius of 0.5~R$_e$. Right-hand panel: $\lambda_{R_e/2}$ vs $\varepsilon_{R_e/2}$. The black dots are the 51 simulated central galaxies from \citet{Martizzi_2014}. The black solid line is the empirical division between fast and slow rotators \citep{Cappellari_2016}.The error bars are the measurement uncertainties propagated from errors on $V$ and $\sigma$.  We find a mean P(SR) = $54\pm7$\% for the whole sample of 22 central galaxies. The mean P(SR) uncertainties are the standard errors on the mean.%
}
\end{center}
\end{figure*}

The $\lambda_R$ profiles of most of the galaxies in \citet{Jimmy_2013} and the new SPIRAL observations do not extend to 1~R$_e$. We therefore select 0.5~R$_e$ as our standard aperture and measure $\lambda_{R_e/2}$. In the cases where the galaxy does not reach 0.5~R$_e$ we use the $\lambda_{R}$ at the maximum radius reached (listed in Table~\ref{tab:kin}). The uncertainties in $\lambda_{R_e/2}$ are propagated using Taylor series expansions from the errors on the $V$ and $\sigma$ measurements.
\\\\
The luminosity-weighted ellipticity, $\varepsilon$, is calculated using the \textsc{idl} routine find\_galaxy.pro (publicly available in the mge\_fit\_sectors package\footnote{http://www-astro.physics.ox.ac.uk/$\sim$mxc/idl/}). This uses the IFS data over the same standard aperture as $\lambda_{R_e/2}$, i.e. $\varepsilon_{R_e/2}$.  
\\\\
With these two parameters we can classify whether the galaxies are slow rotators or not.  A galaxy is defined as slow-rotating if:
\begin{equation}
\lambda_{R_e} < 0.08 + \frac{\varepsilon_{R_e}}{4} \text{ with } \varepsilon_{R_e} < 0.4.
\label{eq:relation}
\end{equation}
\\\\
The relationship from \citet{Cappellari_2016} classifies early-type galaxies into fast and slow rotators from the $\lambda_{R_e}$ and $\varepsilon_{R_e}$ nominally measured at 1~R$_e$, although it is based on ATLAS$^{3D}$ observations that do not always reach 1~R$_e$ \citep{Emsellem_2011, Veale_2016}.  
Our conclusions do not change whether we use the classification for slow rotators based on $\lambda_{R_e/2}$ from \citet[][]{Emsellem_2011} or the \citet[][]{Cappellari_2016} relationship above. We do note that measuring $\lambda$ at radii smaller than $0.5R_e$ does incur an offset to lower values of $\lambda$ \citep[e.g.][]{Veale_2016}. However, from a SAMI stellar kinematic analysis of cluster galaxies (Brough et al., in prep) this would be of the order of  $\Delta \lambda=0.02$ for the sample presented here. This offset is well within the measured uncertainty in $\lambda_{R_e/2}\sim\pm0.07$.
\\\\
In order to test whether the $\lambda_{R_e/2}$ measured from the new SPIRAL observations are directly comparable with the $\lambda_{R_e/2}$ of other central galaxies in the literature, we use the galaxies in \citet[][]{Jimmy_2013}, which were observed with the VIMOS spectrograph and have higher SN, to model the effects of noisier observations. These galaxies are degraded to the same SN as the SPIRAL observations by adding noise to the VIMOS reduced data cubes and running Monte Carlo simulations. We find that noisier observations overestimate the $\lambda_R$ measurements by $0.05\pm0.02$. Therefore, we subtract 0.05 from the SPIRAL $\lambda_R$ profiles (in effect from the $\lambda_{R_e/2}$ measurements). This ensures that the new observations are directly comparable with measurements of other central galaxies in the literature. 
\\\\
The $\lambda_R$ profiles for the newly observed central galaxies and the galaxies in \citet{Jimmy_2013} are shown in the left-hand panel of Figure \ref{fig:rotation}. The final kinematic classification of the galaxies in our sample is based on the empirical relation between $\lambda_{R_e/2}$ and $\varepsilon_{R_e/2}$ (right-hand panel of Figure \ref{fig:rotation}). As the error bars from the $\lambda_{R_e/2}$ measurements are large, and measuring $\lambda$ at radii smaller than $R_e/2$ increases the uncertainty in the measurement, it is possible that a true slow or fast rotator may sit above or below the empirical $\lambda_{R_e/2}$ and $\varepsilon_{R_e/2}$ relation.  We follow \citet[][]{D_Eugenio_2012} and \citet{Houghton_2013} in calculating the probability that each galaxy is a slow rotator by Monte-Carlo modelling $\lambda_{R_e/2}$ over 1000 iterations, assuming a Gaussian error distribution. Galaxies with P(SR)~$>50$\% are classified as slow rotators and the galaxies with P(SR)~$\leq50$\% are classified as fast rotators.  This approach takes into account the limitations of these data, including the radius at which $\lambda$ is measured.
\\\\
The central galaxies in Abell 1689 and Coma have already been assigned a P(SR) by \citet{D_Eugenio_2012} and \citet{Houghton_2013} respectively. For the other galaxies in the literature we maintain their slow/fast classification and calculate their P(SR) using Monte-Carlo modelling as for our observations \citep{Cappellari_2011,Jimmy_2013,Scott_2014,Fogarty_2014}. 
\\\\
Three galaxies in the \citet{Jimmy_2013} sample: 1027, 1042, 2039, have $50<$~P(SR)~$\leq60$\%. This makes their classification uncertain, however, they have been previously classified as slow rotators by \citet{Jimmy_2013}.  We therefore maintain their classification as slow rotators. This does not affect our conclusions which are based on the P(SR) rather than on an absolute slow/fast classification.  
 \\\\
In Table \ref{tab:kin} we show the fiducial radius we reach, $\lambda_{R_e/2}$, $\varepsilon_{R_e/2}$, and the probability of being a slow rotator candidate for each galaxy. 

\begin{table}
\begin{center}
\caption{Central galaxy rotation. The galaxies are sorted by halo mass (see Table \ref{tab:basics}). The top section contains the new SPIRAL observations. The middle section contains the galaxies from \citet{Jimmy_2013}. The bottom section shows other central galaxies from the literature.   Column 2 gives the fiducial radius - the aperture $\lambda_{R_e}$ was measured in (0.5R$_e$ unless otherwise specified). Columns 3 and 4 show the specific angular momentum and ellipticity (see Section 3.5). The $\lambda_{R_e/2}$ uncertainties are propagated from the errors in $V$ and $\sigma$. The $\varepsilon_{R_e/2}$ uncertainties are propagated from the errors in the major and minor axes measurements. Column 5 gives the probability that the galaxy is a slow rotator.}
\begin{tabular}{ l c c c c }
    \hline
    Galaxy & Fiducial & $\lambda_{R_e/2}$ & $\varepsilon_{R_e/2}$ & P(SR) \\ 
     & Radius& $\pm0.070$  & $\pm$0.01 &\%  \\ 
     & [$R_e$] &  &  &  \\ 
    \hline\hline
   2048 & 0.4 & 0.149 & 0.08 &16\\ 
   2121 & 0.5 & 0.164 & 0.20 &24\\ 
   2216 & 0.5 & 0.045 & 0.07 &86 \\ 
   2074 & 0.2 & 0.041 & 0.13 &92\\ 
   2055 & 0.2 & 0.211 & 0.15 &4\\ 
        \hline
    1027& 0.4 &0.106& 0.08&56\\
    1042& 0.4 & 0.085& 0.05 &55\\
    1048& 0.5 & 0.445& 0.33&0\\
    1066& 0.3 & 0.110& 0.35&72\\
     2001&0.5 & 0.124& 0.09&38\\ 
    2086& 0.3 & 0.081& 0.08&63\\
    1261& 0.5 & 0.279& 0.26&1\\
    1050& 0.4 & 0.061& 0.14&80\\
    2039& 0.5 & 0.090& 0.06&53\\
    1153& 0.4 &0.210& 0.28&21\\
     \hline
    12              &0.5& 0.037& 0.15& 95\\
    NGC 4889& 0.5& 0.040& 0.36& 92\\
    019          &0.5& 0.076& 0.24 &95\\
    086          & 1.0& 0.256& 0.45&0\\
    042         & 0.5& 0.043& 0.11&95\\
    M87        & 0.5& 0.019& 0.07&95\\
    NCG 1399 &1.0 &0.080& 0.09&60\\
   
\hline
\end{tabular}
\label{tab:kin}
\end{center}
\end{table}

\section{Results}
\label{sect:results}
We find that the compilation sample of 22 galaxies have mean P(SR) = $54\pm7$\% (errors are standard errors on the mean). 
This percentage is lower than the 70\% found for central galaxies by the previous study of \citet{Jimmy_2013} and could suggest an environmental dependence given the lower halo masses covered by this sample. In the following sections we analyze whether slow rotation in central galaxies is connected to galaxy stellar mass, or environment, or both. 

\begin{figure}
\begin{center}
\includegraphics[width=0.5\columnwidth]{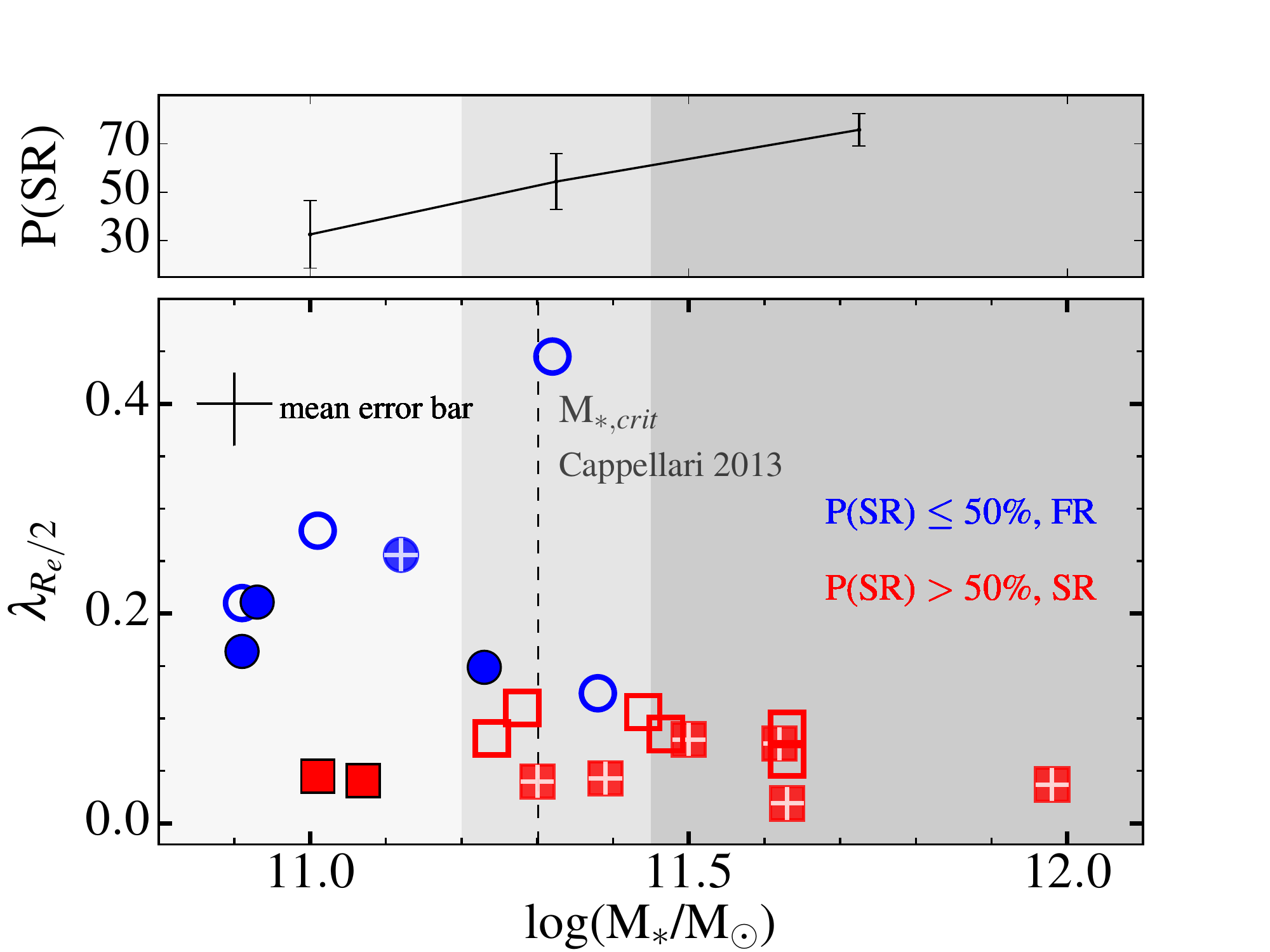}
\caption{\label{fig:L_stellar_mass} 
The lower panel shows $\lambda_{R_e/2}$ as a function of stellar mass. The symbols are color-coded as per Figure \ref{fig:rotation}. The dashed line represents the critical mass from \citet{Cappellari_2013c}. Representative measurement uncertainties are shown at the top of the lower panel. The $\lambda_{R_e/2}$ uncertainties are propagated from errors on $V$ and $\sigma$. The stellar mass uncertainties come from \citet{Taylor_2011}. The upper panel shows the mean P(SR) per stellar mass bin (shaded regions), each stellar mass bin has the same number of galaxies. The mean P(SR) error bars are the standard errors on the mean. The P(SR) increases with increasing stellar mass with a significance of $2.8\sigma$. %
}
\end{center}
\end{figure}

\subsection{Connection between probability of slow rotation and stellar mass}
The rotation of galaxies has been shown to be connected to their stellar mass. Due to their high stellar masses and presumably rich merger histories, the majority of central galaxies are expected to be slow rotators. However, in our sample of central galaxies, with its large stellar mass range, we find that only $54$\% of the galaxies are likely to be slow rotators.  \citet{Cappellari_2013b} and \citet{Cappellari_2013c} showed that in early-type galaxies, around log(M$_{*}/$M$_{\odot}) \sim 11.3$, slow rotators begin to dominate over fast rotators. They defined this as a critical mass (M$_{*, \rm crit}$). However, their sample only includes the central galaxies in the Virgo and Coma clusters and so we examine here how the probability of slow rotation depends on stellar mass for central galaxies.
\\\\
The lower panel of Figure \ref{fig:L_stellar_mass} shows $\lambda_{R_e/2}$ as a function of stellar mass. The sample is divided into three stellar mass bins (shaded regions) with equal number of galaxies per bin. The upper panel shows the mean probability of being a slow rotator candidate per stellar mass bin. The error bars are the errors on the mean. The stellar mass bins and their mean P(SR) are listed in Table \ref{tab:mean}.  We see an increase in mean P(SR) with increasing stellar mass. This result is significant to 2.8$\sigma$.
\\\\

\begin{figure}
\begin{center}
\includegraphics[width=0.5\columnwidth]{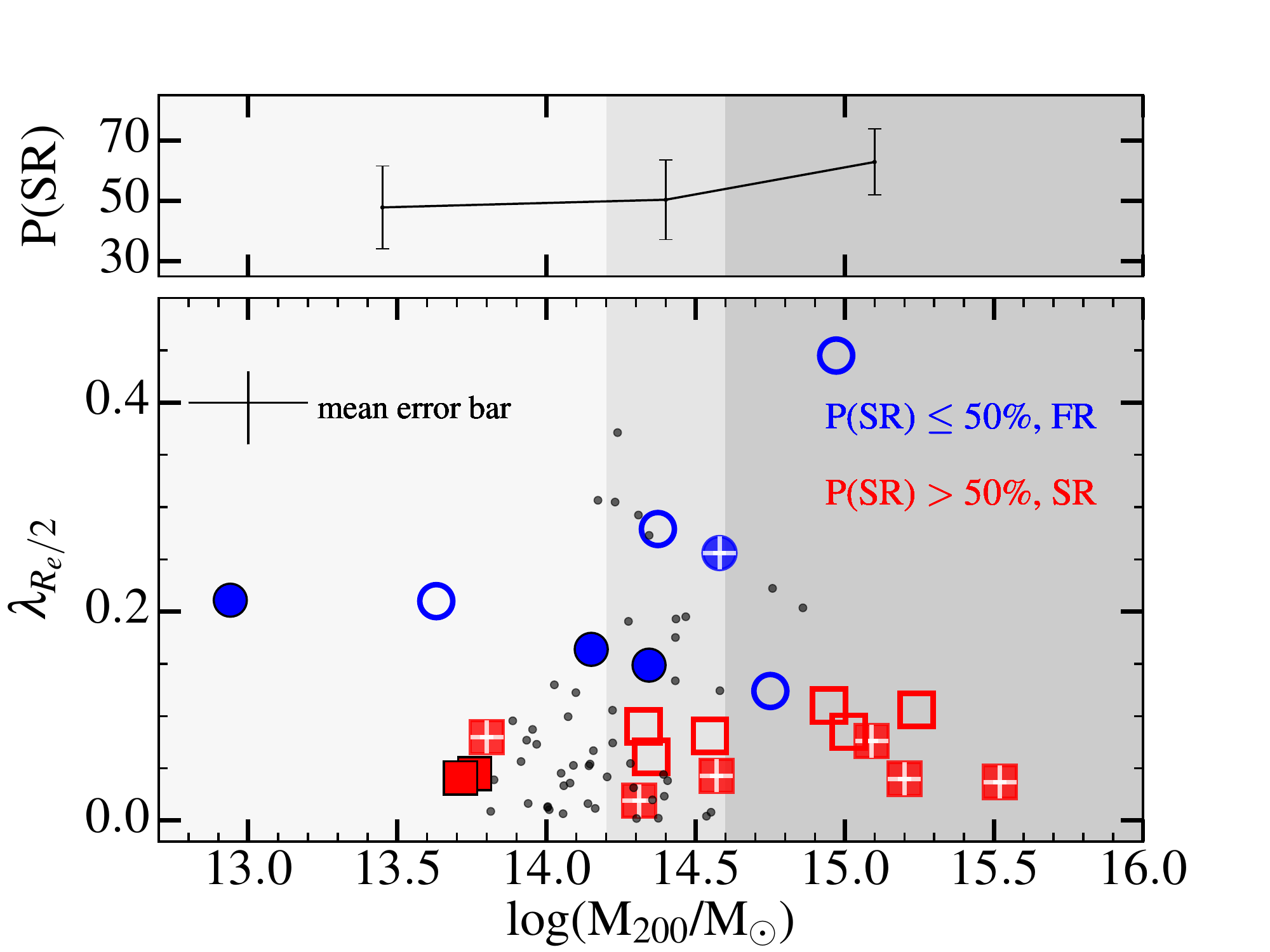}
\caption{\label{fig:m200} The lower panel shows $\lambda_{R_e/2}$ as a function of cluster mass. The symbols and color-coding are as in Figure \ref{fig:rotation}. Representative measurement uncertainties are shown at the top of the lower panel. The $\lambda_{R_e/2}$ uncertainties uncertainties are propagated from errors on $V$ and $\sigma$. The halo mass uncertainties are propagated from $\sigma_{cl}$. The simulations of \citet{Martizzi_2014} are shown as black dots. The upper panel shows the mean P(SR) per halo mass bin (shaded regions), each stellar mass bin has the same number of galaxies. The mean P(SR) error bars are the standard errors on the mean. There is a weak trend of increasing P(SR) with increasing cluster mass, however, it is not statistically significant.%
}
\end{center}
\end{figure}

\subsection{Connection between probability of slow rotation and cluster halo mass}
Due to the small number of central galaxy IFS observations to date, it is not yet known whether central galaxy rotation is influenced by the mass of their host cluster. Here we analyze the effect of environment on the probability of slow rotation. Our sample covers a wide range of group and cluster halo masses, $12.9<$~log(M$_{200}/$M$_{\odot}) <15.6$. 
\\\\
The lower panel of Figure \ref{fig:m200} shows the $\lambda_{R_e/2}$ of central galaxies as a function of their host halo mass. The sample is divided into three halo mass bins (shaded regions) with equal number of galaxies in each bin. The upper panel shows the mean probability of being a slow rotator per halo mass bin. The halo mass bins and their mean P(SR) are listed in Table \ref{tab:mean}.
The P(SR) appears to increase for the massive clusters in our sample, however, this is only significant to 1.8$\sigma$. There is also no sign that the rotation of the 5 galaxies in the low-mass haloes, log(M$_{200}/$M$_{\odot})<14$, is different to those in higher mass haloes.
\\\\
\\\\
The simulations of \citet{Martizzi_2014} consist of 51 clusters with masses $13.8<$~log(M$_{200}/$M$_{\odot})<14.9$. Comparing the 11 observed clusters in the same mass range with their simulations, we find that the mean specific angular momentum measured here ($\lambda_{R_e/2}=0.12\pm0.02$) is consistent with theirs ($\lambda_{R_e}=0.09\pm0.01$). 

\begin{figure}
\begin{center}
\includegraphics[width=0.5\columnwidth]{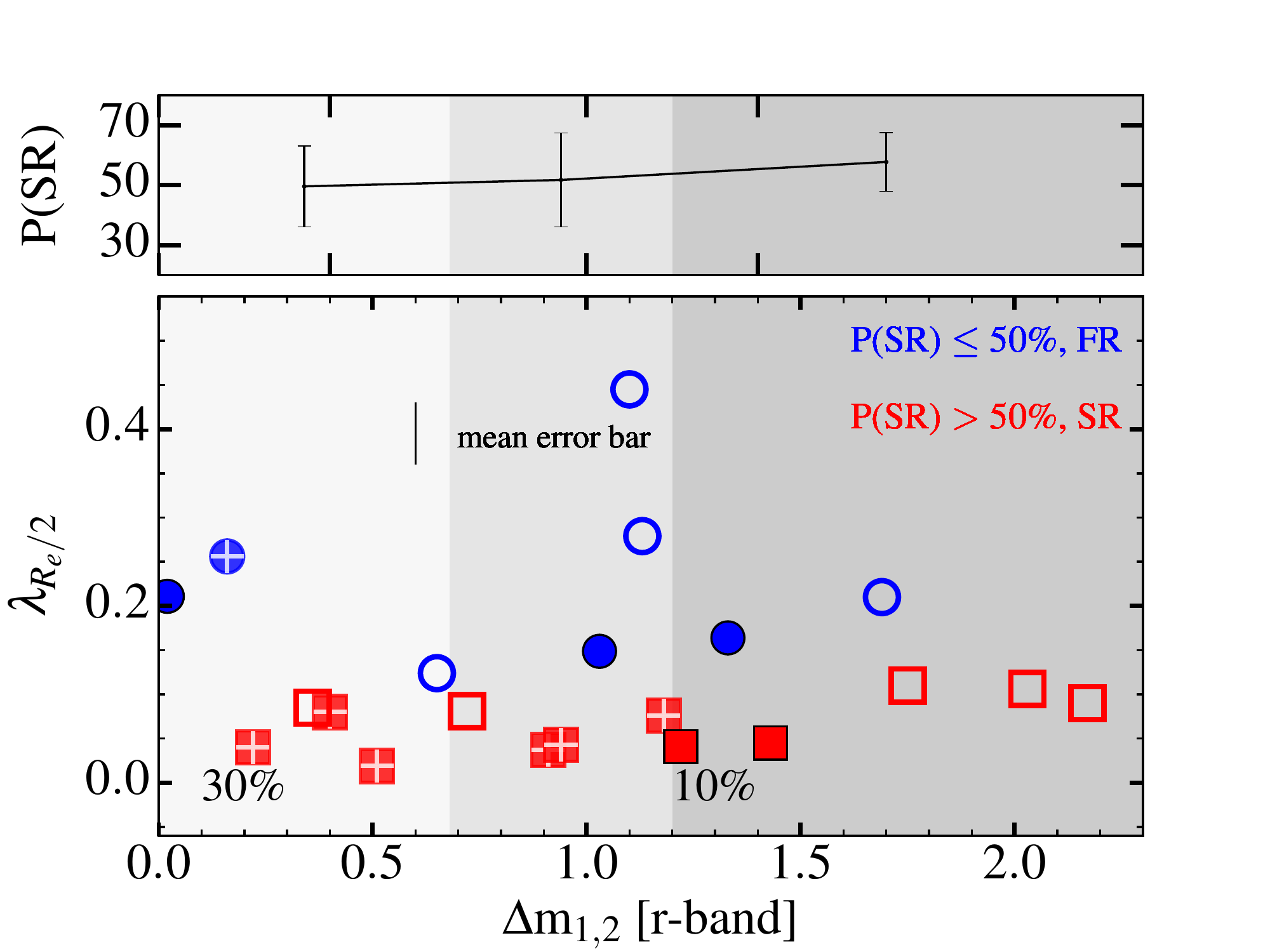}
\caption{\label{fig:dominance}
The lower panel shows $\lambda_{R_e/2}$ as a function of cluster dominance. The symbols are color-coded as per Figure \ref{fig:rotation}. The likely percentages of expected cluster mass residing in cluster substructure at a given dominance are indicated at the bottom of the panel \citep{Smith_2010}. Average uncertainties are propagated from errors on $V$ and $\sigma$. The upper panel shows the mean P(SR) per halo mass bin (shaded regions), each dominance bin has the same number of galaxies. The mean P(SR) error bars are the standard errors on the mean. There is no dependence of P(SR) on cluster dominance.
}
\end{center}
\end{figure}

\subsection{Is the probability of slow rotation connected to the cluster status?}
Dominance is a good indicator of the cluster/group merging status. Small magnitude gaps ($\Delta$m$_{1,2}<1.0$) are likely to indicate a recent halo merger. This is supported by the high percentage of the mass of these clusters residing in cluster substructure \citep[$\sim 30$\%;][]{Smith_2010}. 
\\\\
Figure \ref{fig:dominance} explores the connection between the probability of slow rotation and cluster dominance. The lower panel shows $\lambda_{R_e/2}$ as a function of cluster dominance, and the upper panel shows the P(SR) per dominance bin (shaded regions). Each bin contains the same number of galaxies. The dominance bins and their mean P(SR) are listed in Table \ref{tab:mean}.  
We find no significant dependence of P(SR) on cluster dominance, with a difference from low to high dominance of $0.9\sigma$.

\begin{table}
\begin{center}
\caption{Mean probabilities of slow rotation per stellar mass, cluster mass and cluster dominance. Each bin contains an equal number of galaxies.}
\begin{tabular}{ l l c c }
    \hline
     &bin size &mean & error on \\ 
     &&P(SR) [\%]& the mean\\
      \hline\hline
      Stellar mass    &10.90 - 11.20& 32.6&13.9\\
      log(M$_\odot$)&11.20 - 11.45&54.4&11.1\\
      &11.45 - 12.0& 75.7& 6.7\\
      \hline
      Cluster halo mass&12.70 - 14.20&47.8&13.7\\
      log(M$_\odot$)&14.20 - 14.60 &50.4&13.2\\
      &14.60 - 15.60&62.9&11.0\\
      \hline
      Cluster dominance&0.00 - 0.68 &49.6& 13.5\\
      $\Delta$m$_{1,2}$&0.68 - 1.20 &51.7&15.7\\
      $[r-$band$]$&1.20 - 2.30 &57.7&9.8\\
           
    \hline
\end{tabular}
\label{tab:mean}
\end{center}
\end{table}

\section{Discussion} 
\label{sec:d}
We have presented the specific angular momentum of central galaxies from IFS observations. We include new observations of 5 central galaxies observed with SPIRAL on the AAT at redshifts $0<z<0.1$. These galaxies, together with the 10 central galaxies presented in \citet{Jimmy_2013}, and the central galaxies in the Abell 85, 168, 1689 and 2399, Coma, Fornax and Virgo clusters, span a wide range of halo masses, $12.9<$~log(M$_{200}/$M$_{\odot})<15.6$, as well as a wide range of stellar masses, $10.9<$~log(M$_{*}/$M$_{\odot})<12.0$. The composite sample presented here is not complete, however, it is representative of a wide range of environments. This is the first analysis of the role of environment on the stellar kinematics of central galaxies using IFS.
\\\\
The connection between stellar mass and galaxy rotation was previously explored for the early-type galaxy population by the ATLAS$^{3D}$ \citep{Cappellari_2013c} and MASSIVE teams \citep{Veale_2016}. We find a similar trend in central galaxies to that observed for the general and massive early-type galaxy populations: the probability of slow rotation increases with increasing galaxy stellar mass. Above the ATLAS$^{3D}$ critical stellar mass \citep[log(M$_{*, \rm crit}/$M$_{\odot})=11.3$][]{Cappellari_2013c} the probability of slow rotation in central galaxies, P(SR)~$=68\pm8\%$, is higher than in lower stellar mass galaxies, P(SR)~$=38\pm10\%$ (Figure~\ref{fig:L_stellar_mass}). 
\\\\
Central galaxy stellar mass growth is predicted to be directly influenced by their host cluster halo mass growth \citep{White_1978,Khochfar_2003,De_Lucia_2007,Oser_2010,De_Lucia_2012} and more massive cluster halos tend to host more massive central galaxies \citep[e.g.][]{Brough_2008,Lidman_2012,Stott_2012,Burke_2013,Oliva_Altamirano_2014,Luparello_2015,Shankar_2015}. To explore this further we examined the connection between the cluster mass and P(SR) for the galaxies in our sample.  We do not find a significant dependence of the probability of central galaxy slow rotation on host cluster mass (Figure~\ref{fig:m200}).  
\\\\
We use the cluster dominance as a proxy for cluster-cluster mergers to explore whether central galaxy slow rotation is influenced by cluster-scale effects. Clusters with low dominance ($\Delta$m$_{1,2} <1$) have been predicted to have gone through recent cluster mergers \citep[][]{Dariush_2010, Smith_2010}. These mergers would enhance the probability of galaxy-galaxy major mergers \citep[see][]{Martel_2014}. This could increase the angular momentum in central galaxies. A cluster with no recent interactions, on the contrary,  tends to have a well-established central galaxy, which is more likely to experience minor mergers than major mergers. These clusters are gas depleted which causes a spin-down in the central galaxy. Minor mergers are not powerful enough to change the spin in central galaxies, therefore, they generate slow-rotating galaxies \citep[][]{Naab_2014}.
\\\\
When we compare the dependence of the P(SR) on the dominance of all 22 clusters in our sample, we do not find a trend. Central galaxies with a high probability of slow rotation are hosted by clusters with both low and high dominance. 
This suggests that cluster mergers do not play a significant role in central galaxy rotation (Figure~\ref{fig:dominance}). 
\\\\
In order to further probe the influence of cluster evolution on the angular momentum of central galaxies, new cosmological simulations encompassing the whole group-cluster mass range are needed. However, central galaxies represent a challenge for cosmological simulations. \citet{Martizzi_2014} showed the practical difficulties of modelling such complex galaxies. Their massive halos require bigger cosmological boxes than those currently available, and the implementation of many physical processes acting together at a given time. While the implementation of AGN feedback has brought simulated galaxies into agreement with observations, the limited halo mass range simulated at the moment restricts further comparison with these observations.
\\\\
In summary, we present for the first time, a study of the probability of slow rotation in central galaxies and its connection with their host environment. We find that the probability of slow rotation in central galaxies increases as the central galaxy increases in stellar mass. Our observations are consistent with the trend with stellar mass seen in early-type galaxies by \citet{Cappellari_2013c} and \citet{Veale_2016}.  However, we do not observe a significant dependence of the probability of slow rotation in central galaxies on environment as described by their host cluster halo mass or dominance.  This result is in tension with models that use halo mass as the dominant predictor of central galaxy properties \citep[e.g.][]{Behroozi_2010, Mutch_2013, Moster_2013}.  
Larger samples and more detailed simulations are needed to further explore the weak trend with cluster halo mass that we observe here.

\section{Conclusions}
\label{sect:conclusions}
We present a pilot analysis of the role of environment on the rotation of 22 central galaxies from a composite sample of new (5 galaxies) and previous IFS observations (17 galaxies).  To ensure a fair comparison across the heterogeneous sample we have paid particular attention to measuring the key galaxy properties uniformly across the 22 galaxies.  The sample spans a cluster halo mass range of $12.9<$~log(M$_{200}/$M$_{\odot})<15.6$ and a stellar mass range of  $10.9<$~log(M$_{*}/$M$_{\odot})<12.0$. 
\\\\
To analyse the correlation between galaxy rotation and stellar mass, cluster halo mass, and cluster dominance, we compute a probability of slow rotation for each galaxy in our sample. This probability is calculated from the position of the galaxy in the $\lambda_{R_e/2}$ - $\varepsilon_{R_e/2}$ diagram 
over 1000 Monte Carlo simulations.
\\\\
Our results suggest a connection between the probability that a central galaxy is a slow rotator and its stellar mass. The probability of slow rotation is higher in the most massive central galaxies.  However, when we examine the dependence of slow rotation on host cluster halo mass we do not observe a significant relationship. We also find that cluster dominance has no significant effect on the probability of slow rotation in central galaxies.  This is in contradiction with models that use halo mass alone to predetermine central galaxy properties.
\\\\
The next generation of IFS surveys will bring significantly increased sample sizes: MaNGA \citep[][]{Bundy_2014}, SAMI \citep[][]{Bryant_2015} and HECTOR \citep[][]{HECTOR}; offer the chance to substantiate the findings presented here.

\section{Acknowledgements}
We thank the anonymous referee for their constructive comments that have helped to improve this paper.
\\\\
POA acknowledges the valuable feedback of Chris Lidman, Luca Cortese, and Edoardo Tescari, as well as the ARC Discovery Project DP130101460.
\\\\
SB acknowledges funding support from the Australian Research Council through a Future Fellowship (FT140101166)
\\\\
MALL acknowledges support from UNAM through the PAPIIT project IA101315
\\\\
Funding for the Sloan Digital Sky Survey (SDSS) has been provided by the Alfred P. Sloan Foundation, the Participating Institutions, the National Aeronautics and Space Administration, the National Science Foundation, the U.S. Department of Energy, the Japanese Monbukagakusho, and the Max Planck Society. The SDSS Web site is http://www.sdss.org/.
\\\\
The SDSS is managed by the Astrophysical Research Consortium (ARC) for the Participating Institutions. The Participating Institutions are The University of Chicago, Fermilab, the Institute for Advanced Study, the Japan Participation Group, The Johns Hopkins University, Los Alamos National Laboratory, the Max-Planck-Institute for Astronomy (MPIA), the Max-Planck-Institute for Astrophysics (MPA), New Mexico State University, University of Pittsburgh, Princeton University, the United States Naval Observatory, and the University of Washington.
\\\\
 This research has made use of the NASA/IPAC Extragalactic Database (NED) which is operated by the Jet Propulsion Laboratory, California Institute of Technology, under contract with the National Aeronautics and Space Administration. 

\end{document}